\documentclass[preprint,12pt]{elsarticle}
\usepackage{graphicx}
\usepackage{enumerate}
\usepackage{amsmath}
 \usepackage{framed}
  \usepackage{setspace}
  \usepackage{amssymb}
  \usepackage[justification=centering]{caption}
\usepackage{algorithm}
\usepackage{algpseudocode}
\usepackage{comment}
\usepackage{bm}
\usepackage{multirow}%
\usepackage{amsfonts}%
\usepackage{amsthm}%
\usepackage{mathrsfs}%
\usepackage[title]{appendix}%
\usepackage{xcolor}%
\usepackage{textcomp}%
\usepackage{manyfoot}%
\usepackage{booktabs}%
\usepackage{algorithm}%
\usepackage{algorithmicx}%
\usepackage{algpseudocode}%
\usepackage{listings}%

\makeatletter
\renewcommand{\ALG@beginalgorithmic}{\normalsize}
\makeatother

\usepackage{float}
\usepackage{lipsum}



\begin{document}

\begin{frontmatter}
\title{Quantum Circuit Implementation and Resource Analysis of LBlock and LiCi}


%\author[label1]{Li Yang\corref{1}}\ead{yang@is.ac.cn}
%\cortext[1]{Corresponding author.}
%\ead{yang@is.ac.cn}
\author{XiaoYu Jing$^{1}$}
\author{YanJu Li$^{1,2}$}%\ead{yangli@iie.ac.cn}
\author{GuangYue Zhao$^1$}
\author{Huiqin Xie$^1$\corref{1}}
\cortext[1]{Corresponding author email: xiehuiqindky@163.com}
\address{1.Beijing Electronic Science and Technology Institute, Beijing 100070, China\\
2.Guangxi Key Laboratory of Cryptography and Information Security, Guilin University of Electronic Technology, Guilin {\rm 541004}, China}

\begin{abstract}
Due to Grover's algorithm, any exhaustive search attack of block ciphers can achieve a quadratic speed-up. To implement Grover,s exhaustive search and accurately estimate the required resources, one needs to implement the target ciphers as quantum circuits. Recently, there has been increasing interest in quantum circuits implementing lightweight ciphers. In this paper we present the quantum implementations and resource estimates of the lightweight ciphers LBlock and LiCi. We optimize the quantum circuit implementations in the number of gates, required qubits and the circuit depth, and simulate the quantum circuits on ProjectQ. Furthermore, based on the quantum implementations, we analyze the resources required for exhaustive key search attacks of LBlock and LiCi with Grover's algorithm. Finally, we compare the resources for implementing LBlock and LiCi with those of other lightweight ciphers.

\end{abstract}


\end{frontmatter}

% \linenumbers

\section{Introduction}
The rapid development of quantum computing is threatening the security of classical cryptosystems. Due to Shor's algorithm, many widely deployed public-key algorithms, such as RSA, ECC, and DSA, will be broken once large-scale quantum computers are built. 
	
	In the field of symmetric cryptography, the research on the impact of quantum algorithms on symmetric-key ciphers has also received substantial attention in recent years. The application of Grover's algorithm \cite{bib1} can provide a quadratic speedup for any exhaustive search attack. If the keyed quantum encryption oracle is available, an attacker with a quantum computer can break many symmetricschemes by Simon's period-finding algorithm \cite{bib2,bib3}. Considering the rapid development of quantum computers, it is urgent to reevaluate the security of symmetric ciphers in quantum computing environment. 
	
	In 2010, based on Simon's algorithm \cite{bib4}, Kuwakado et al. effectively distinguished 3-round Feistel structure from random permutations\cite{bib5}. In 2012, Kuwakado used the Simon's algorithm for Even-Mansour scheme \cite{bib6} and reduced the time complexity of key recovery to polynomial time. In 2014, Kaplen et al. proposed a Quantum version \cite{bib9} of the classical meet-in the-middle attack \cite{bib8} using the quantum walk algorithm proposed by Ambainis \cite{bib7}. In 2015, Roetteler and Steinwandt \cite{bib10} proposed that,related-key attacks can be reduced to polynomial time using Simon's algorithm, when combined with the ability to make quantum superposed queries.In 2016, Santoli and Schaffner \cite{bib11} used Simon's algorithm to attack symmetric-key cryptographic primitives. In 2017, Leander and May \cite{bib12}  combined Simon's algorithm with Grover's search algorithm to implement a quantum key-recovery attack on FX structure. In the same year, Hosoyamada and Aoki \cite{bib13} extend Simon's quantum algorithm so that we can recover the hidden period of a function that is periodic only up to constant.In 2018, DONG and WANG \cite{bib14} used Grover's and Simon's algorithms to generate new quantum key-recovery attacks on different rounds of Feistel constructions.In 2019, Xie proposed a quantum distinguisher 3-round Feistel structure and Even-Mansour cipher based on the Bernstein-Vazirani algorithm [15]. In the same year, DONG et al \cite{bib16}. studied the quantum distinguishers about some generalized Feistel schemes.In 2021, Zhou combined the distinguisher proposed in \cite{bib15} with Grover's algorithm to generate a quantum key-recovery attack on different rounds of Feistel constructions \cite{bib17}.
	
	To accurately evaluate the quantum security of block ciphers, we need to analyze the resources consumed by various quantum attacks on the block ciphers. More resources the attacks require, the more secure the block ciphers are in the post-quantum era. Quantum attacks on a block cipher often necessitate the implementation of its quantum circuit. A typical example is the exhaustive key search attack, which requires iterative execution of the unitary operator corresponding to the encryption algorithm. Therefore, the quantum circuit implementations of block ciphers and their optimization are the premise of the resources analysis of various quantum attacks and, subsequently, the foundation of evaluating quantum security of block ciphers. In this field, researchers first focused on the quantum circuit implementations of AES. In 2016, Grassl et al. \cite{bib18} proposed a quantum circuit to implement AES. They analyzed the resources required for the performance from three perspectives: quantum gates, circuit depth and qubits. Later, Kim et al. \cite{bib19} improved the byte substitution operation by saving a multiplication operation based on the work of Grassl et al \cite{bib18}. In 2020, Langenberg et al. designed a quantum circuit for byte substitution \cite{bib21} based on Boyar et al.'s classical algorithm \cite{bib20}, which reduced the number of Toffoli gates by 88$\%$ compared to Grassl et al.'s scheme.
	
	Post-quantum security for lightweight block ciphers has recently been a popular research direction. Anand et al. \cite{bib22} studied the resource requirements of the SPECK algorithm for quantum key search under the model of known plaintext attack and provided the results of quantum differential cryptanalysis. He proposed the optimal quantum circuit for Simon's algorithm in \cite{bib23} and simulated its implementation on the Qiskit. Jang et al. applied LIGHTER-R to optimize the quantum circuit of Sboxes of PRESENT and GIFT \cite{bib24} algorithms, which thereby offered quantum circuit implementation for algorithms. In addition, he conducted quantum circuit analysis for HIGHT, CHAM, LEA SPECK \cite{bib25} and default algorithms \cite{bib26}, and simulated the quantum circuit implementation of the default algorithm in ProjectQ \cite{bib27}.

	In this paper, we propose the quantum circuit implementations for lightweight block ciphers LBlock and LiCi. We implement the Sboxes of the algorithm at a small cost and try to use the least quantum resources in the overall circuit implementations. We simulated our schemes in ProjectQ and analyzed the resources required by Grover's exhaustive search attack on LBlock and LiCi .
	
	The contributions of this paper are as follows:
	
1. We propose the in-place quantum circuit implementations of LBlock and LiCi and analyze their quantum security.
	
	2. In our circuit implementations, we optimize the resource consumption of each component from the perspectives of quantum gates, qubits and circuit depth. We first optimize the implementations of the non-linear component S-boxes so that they consume as few quantum gates as possible without using auxiliary qubits. Afterwards, we focus on the overall structure. Observing that the last Toffoli gate in the quantum circuit of each S-box of LBlock can be derictly applied to the right branch by taking the qubits in the right branch as targer qubits, we reduce one Toffoli gate for the implemetation of each S-box in the inversion process, making the whole circuit reduce 256 Toffoli gates.
	
	3. In the post-quantum era, the more quantum cost required for executing attacks, the more secure the block ciphers are. We compare the required resources of quantum implementations of LBlock, LiCi with other lightweight block ciphers. We also compare the quantum security of LBlock and LiCi with the levels of quantum security specified by the United States? National Institute of Standards and Technology (NIST). 
	
The remainder of this paper is organized as follows. Section 2 introduces the lightweight block ciphers LBlock and LiCi, Grover's algorithm, basic quantum gates, LIGHTER-R technique and ProjectQ. In Section 3 and Section 4 we present the quantum circuit implementations in detail, including the round function and the key scheduling, and analyze the resources of the schemes. Section 5 explores the resources required for quantum key recovery for LBlock and LiCi. Finally, the whole paper is concluded with a summary and an outlook.

\section{Preliminary}
\subsection{Quantum Gates}
Several commonly used quantum gates implement quantum circuits 
	for block ciphers, including X, CNOT, and Toffoli. The function 
	of the X gates is equivalent to the NOT operation in classical 
	circuits: the inverse of the input, i.e., $X(a)=\sim a$, as in 
	Figure \ref{fig:fig1}(A).The CNOT gate is similar to the XOR operation in classical 
	circuits. It has two qubits as the input, including a control qubit 
	and a target qubit. The target qubit will be flipped if the value of 
	the control qubit is 1, while it does not change when the value of 
	the control qubit is 0, i.e., $CNOT(a,b)=(a,a\oplus b)$, as in 
	Figure \ref{fig:fig1}(B). Swapping the two qubits can be realized by Swap gates. 
	A Swap gate consists of three CNOT gates, as shown in Figure \ref{fig:fig1}(C). 
	The Toffoli gates receive three qubits as the input, two of which 
	are control qubits, and the third is the target qubit. The target 
	qubit is flipped when the value of both control qubits is 1, i.e., 
	$Toffoli(a,b,c)=(a,b,ab\oplus c)$, as in Figure \ref{fig:fig1}(D). Concretely, a 
	Toffoli gate consists of Clifford gates(CNOT gates, H gates, and P 
	gates) and a T gate. Using the scheme of \cite{bib28}, a Toffoli gate can 
	be decomposed into 7 T gates and 8 Clifford gates with a T depth 
	of 4 and a total depth of 8, as shown in Figure \ref{fig:fig2}.
	
	\begin{figure}
		\centering
		\includegraphics[width=10cm]{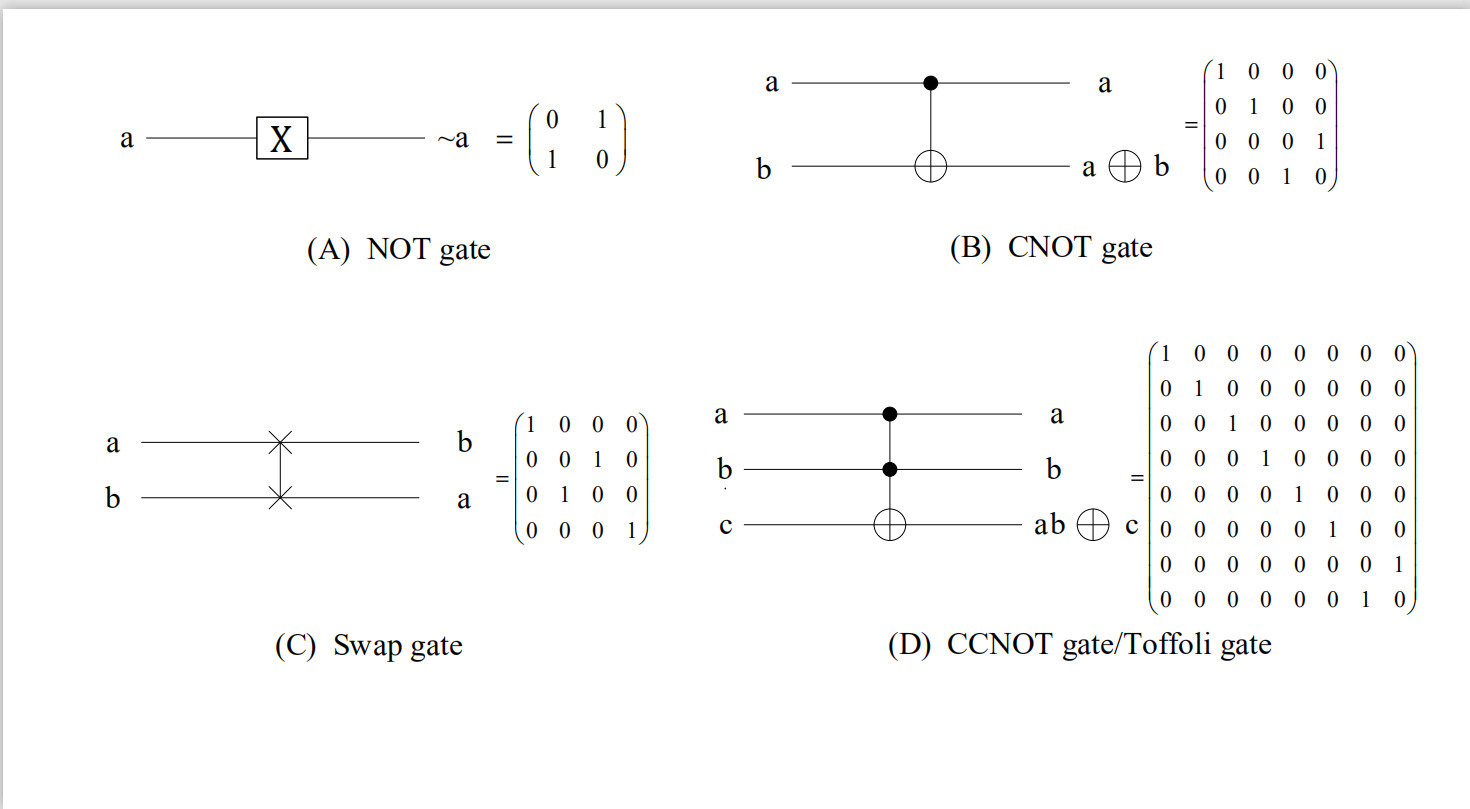}\\
		\caption{Quantum gates required for quantum circuit implementation}
	\end{figure}

	\begin{figure}[!h]
		\centering
		\includegraphics[width=5in]{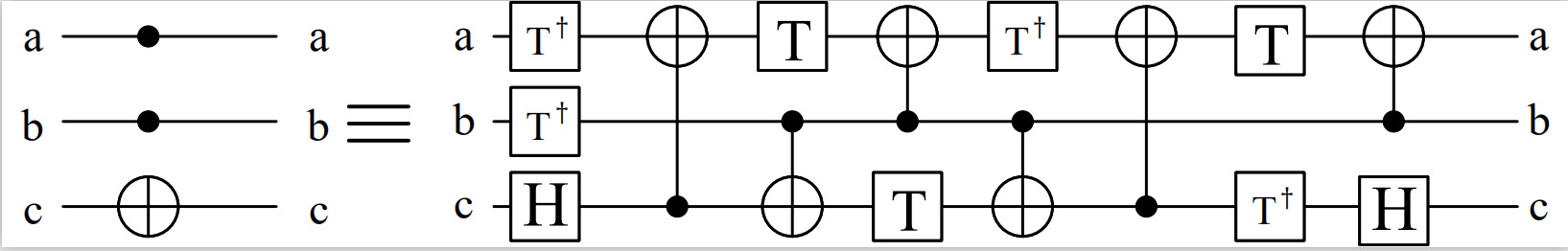}
		\caption{In-place implementation of Toffoli gates using H gates, T gates, and CNOT gates}\label{fig:fig2}
	\end{figure}
	
	The H gates, T gates, T$^{\dag}$ gates used in the decomposition process of Toffoli gates are defined as follows:
	\begin{equation}\label{eq:eqution1}
		H = \frac{1}{\sqrt 2}
		\begin{pmatrix}
			1&1\\1&-1
		\end{pmatrix},
		T=
		\begin{pmatrix}
			1&0\\0&\frac{1+i}{\sqrt{2}}
		\end{pmatrix}
		,T^{\dag}=
		\begin{pmatrix}
			1&0\\0&\frac{1-i}{\sqrt{2}}
		\end{pmatrix}
	\end{equation}
	
	Same as in \cite{bib18}, we generally do not distinguish between T gates and T$^{\dag}$gates, referring to them collectively as T gates.

\subsection{Grover algorithm}\label{sec3}
	Suppose there is a search problem that we need to find a particular element in N unsorted data, where $N=2^N$. Define a Boolean function $f$ on $\{0,1\}^n \to \{0,1\}$ that represents the elements in the data set as integers from 0 to N-1. If $x$ is the element we are looking for, then $f(x)=1$, otherwise $f(x)=0$.
	
	The Grover \cite{bib1} algorithm proceeds as follows:
	
	(1) Initialize n qubits $\lvert 00...0 \rangle$. 
	
	(2) Apply Hadamard transformation on $\lvert 00...0\rangle$:
	
	\begin{equation}\label{eq:equation2}
		\lvert \psi \rangle = \frac{1}{\sqrt{N}} \sum_{x \in \{0,1\}^n} \lvert x \rangle
	\end{equation}
	
	(3) According to \cite{bib29}, execute Grover iteration $\lfloor \frac{\pi}{4} \sqrt{N} \rfloor$ times. Grover's iteration consists of two parts: the oracle and the diffusion operator:
	
	Oracle: apply oracle to the state $\lvert x \rangle \lvert q \rangle$, where $\lvert x \rangle$  are n qubits and $\lvert q \rangle$ is a single qubit. The function of an oracle is denoted as $\lvert x \rangle \lvert q \rangle \to \lvert x \rangle \lvert q \oplus f(q) \rangle$. If the initial value of $\lvert q \rangle$ is $(\lvert 0 \rangle -\lvert 1 \rangle / \sqrt{2})$, the action can be notated as $\lvert x\rangle (\lvert 0\rangle -\lvert 1\rangle / \sqrt{2})\xrightarrow{O} (-1)^{f(x)} \lvert x\rangle (\lvert 0\rangle -\lvert 1\rangle / \sqrt{2}) $and can be further simplified to $\lvert x\rangle \xrightarrow{O} (-1)^{f(x)} \lvert x\rangle$. Consequently, oracle marks the solutions to the search problem by shifting the phase of the solution.
	
	Diffusion operator: amplitude the amplitude of the solution by performing the diffusion operator.
	
	(4) Measure the result, able to find the target element with a probability of close to 1.
	
	Through the above analysis, when we use the Grover algorithm to attack block cipher, the time complexity of the Grover quantum search is $O(2^{\frac{n}{2}})$ while that of the classical algorithm is $O(2^n)$, which achieves a square-level acceleration with a security effect equivalent to halving the algorithm's key length.

	\begin{figure}
		\centering
		\includegraphics[width=5in]{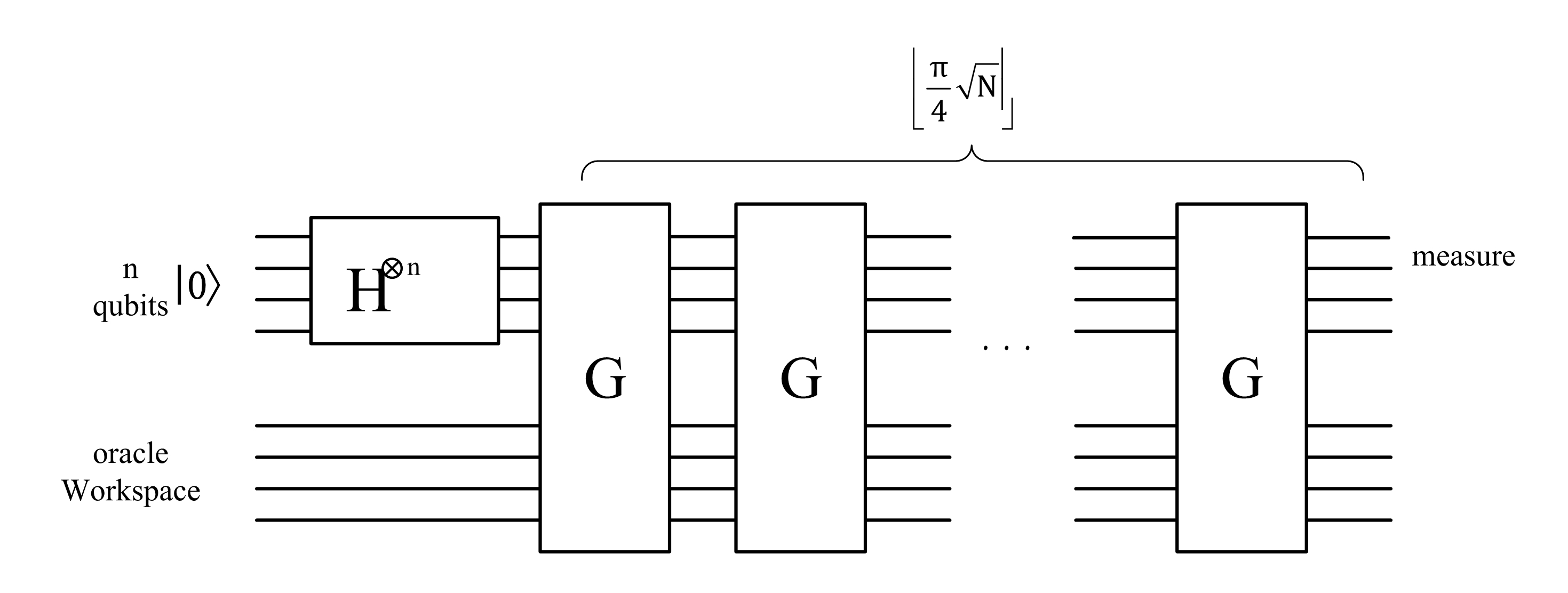}
		\caption{Grover quantum search}\label{fig:fig3}
	\end{figure}
	\subsection{LIGHTER-R}\label{subsec3}
	In \cite{bib30}, Jean introduces a graph-based Meet In The Middle (MITM) 
	approach that gets a compact implementation of lightweight 
	encryption building blocks given specific weighted instructions set and their related costs. In addition, the authors introduce another algorithm that uses 
	the partitioning theory to achieve a trade-off between the 
	output's optimality and the computation's tractability. 
	Based on these two algorithms, the authors propose an 
	automated tool called LIGHTER, which can be used to find 
	optimized implementations of lightweight basic building blocks.
	
	Based on Jean's work, Dasu in \cite{bib31} proposed a circuit implementation tool 
	for a 4-bit Sbox in quantum computers, namely LIGHTER-R. LIGHTER-R can be 
	considered an extension of LIGHTER, and it is possible 
	to provide optimized quantum circuits for the 4-bit Sbox 
	based on reversible logic libraries (e.g., MCT, NCT).
	
	\subsection{ProjectQ}
	ProjectQ \cite{bib27} is an open-source software framework for quantum computing implemented in Python. It can realize the simulation of quantum circuits, and our resource estimation of the quantum circuit is based on this framework.
	
	\subsection{LBlock}
	
\begin{figure}[!h]
		\centering
		\includegraphics[width=2in]{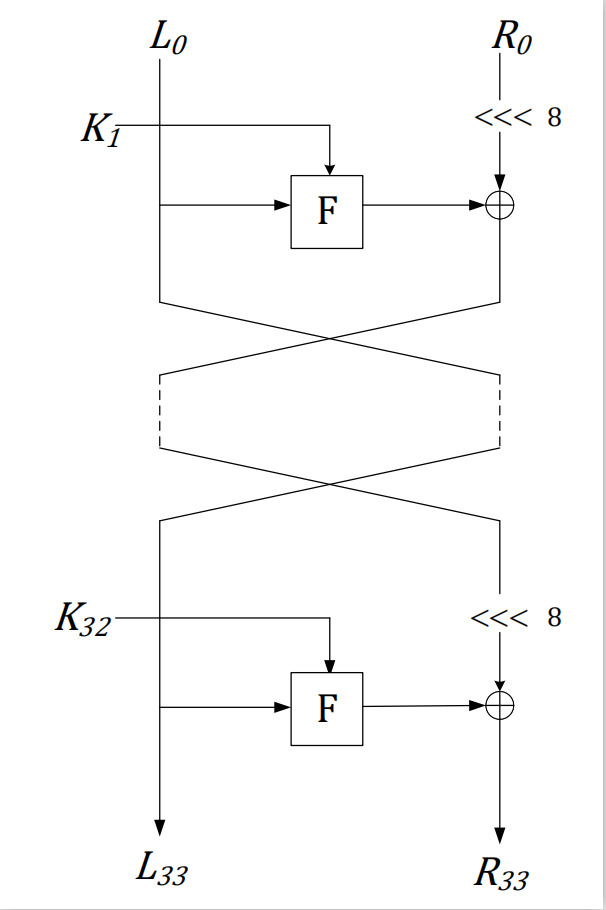}
		\caption{Encryption procedure of LBlock}\label{fig:fig4}
	\end{figure}

	The LBlock \cite{bib32} is a lightweight block cipher proposed by Wu et al. at the ACNS in 2011. The block length of LBlock is 64 bits, and the key size is 80 bits. It employs a variant Feistel structure and consists of 32 rounds. Because of its efficient implementation and good application performance on hardware and software platforms, the algorithm can be applied in resource-constrained environments such as RFID. The plaintext is denoted as $M=L_0 \| R_0$while the output $C=L_{33} \| R_{33}$ as the 64-bit ciphertext. The encryption process is shown in Figure \ref{fig:fig4}.
	
	\subsubsection{Round Function}
	
	The round function consists of AddRoundKey, non-linear layers, and diffusion layers. The non-linear layer consists of eight 4-bit S-boxes $s_{i} (0\leq i\leq 7)$ in parallel. The diffusion layer is defined as a permutation of eight 4-bit words. The expression is as follows:
	
	\begin{equation}\begin{split}\label{eq:equation3}
			Z=Z_{7}\|Z_{6}\|Z_{5}\|Z_{4}\| Z_3\|Z_2\|Z_1\|Z_0 \\ \implies Z^{'} =Z_6\|Z_4\|Z_7\|Z_5\|Z_2\|Z_0\|Z_3\|Z_1
		\end{split}
	\end{equation}
	
\begin{figure}[h]
		\centering
		\includegraphics[width=4in]{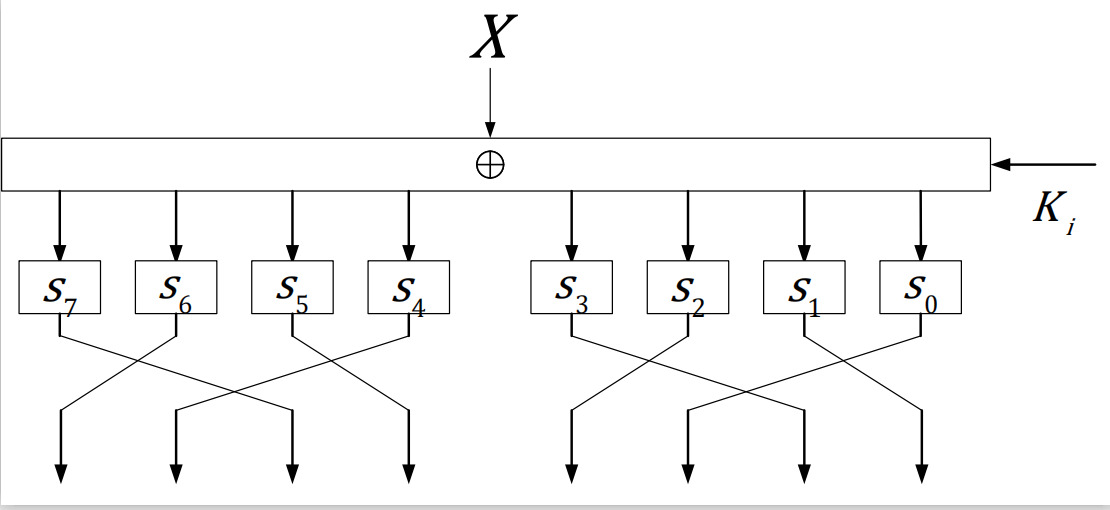}
		\caption{5 Round function of LBlock}\label{fig:fig5}
	\end{figure}

	The structure of round function F is shown in Figure \ref{fig:fig5}.
	
	\subsubsection{Key Scheduling}
	
	For the 80-bit master key $K$, the leftmost 32 bits of $K$ are the subkey $K_1$, and then update the round key for the ith round of output as follows:
	
	(1)$K<<<29$;
	
	(2)$[k_{79}k_{78}k_{77}k_{76}]=s_{9}(k_{79}k_{78}k_{77}k_{76}),[k_{75} k_{74} k_{73} k_{72}]=s_{8}(k_{75} k_{74} k_{73} k_{72})$;
	
	(3)$[k_{50}k_{49}k_{48}k_{47}k_{46}]=[k_{50}k_{49}k_{48}k_{47}k_{46}]\oplus [i]_2$
	
	where $s_9$  and $s_8$ are two 4-bit S-boxes, and the S-boxes used in the round function and key scheduling are shown in Table \ref{tab:table1}.
	
	\begin{table*}
		\centering
		\caption{S-boxes used in LBlock}
		\label{tab:table1}
		\begin{tabular}[htbp]{|p{0.3em}|p{0.2em}|p{0.2em}|p{0.2em}|p{0.2em}|p{0.2em}|p{0.2em}|p{0.2em}|p{0.2em}
				|p{0.2em}|p{0.2em}|p{0.2em}|p{0.2em}|p{0.2em}|p{0.2em}|p{0.2em}|p{0.2em}|}
			
			\hline
			x&0&1&2&3&4&5&6&7&8&9&a&b&c&d&e&f\\ \hline
			$s_0$&e&9&f&0&d&4&a&b&1&2&8&3&7&6&c&5\\ 
			\hline
			$s_1$&4&b&e&9&f&d&0&a&7&c&5&6&2&8&1&3\\ 
			\hline
			$s_2$&1&e&7&c&f&d&0&6&b&5&9&3&2&4&8&a\\ 
			\hline
			$s_3$&7&6&8&b&0&f&3&e&9&a&c&d&5&2&4&1\\ \hline
			$s_4$&e&5&f&0&7&2&c&d&1&8&4&9&b&a&6&3\\ \hline
			$s_5$&2&d&b&c&f&e&0&9&7&a&6&3&1&8&4&5\\ \hline
			$s_6$&b&9&4&e&0&f&a&d&6&c&5&7&3&8&1&2\\ \hline
			$s_7$&d&a&f&0&e&4&9&b&2&1&8&3&7&5&c&6\\ \hline
			$s_8$&8&7&e&5&f&d&0&6&b&c&9&a&2&4&1&3\\ \hline
			$s_9$&b&5&f&0&7&2&9&d&4&8&1&c&e&a&3&6\\ \hline
			
		\end{tabular}
		
	\end{table*}

	\subsection{LiCi}
	LiCi is a lightweight block cipher proposed by Patil et al. in 2017 \cite{bib33}. The block length of LiCi is 64-bit, and the key size is 128-bit. It is a balanced Feistel structure network and has 31 rounds.

	\subsubsection{Encryption procedure}
	
\begin{figure}[!h]
		\centering
		\includegraphics[width=3in]{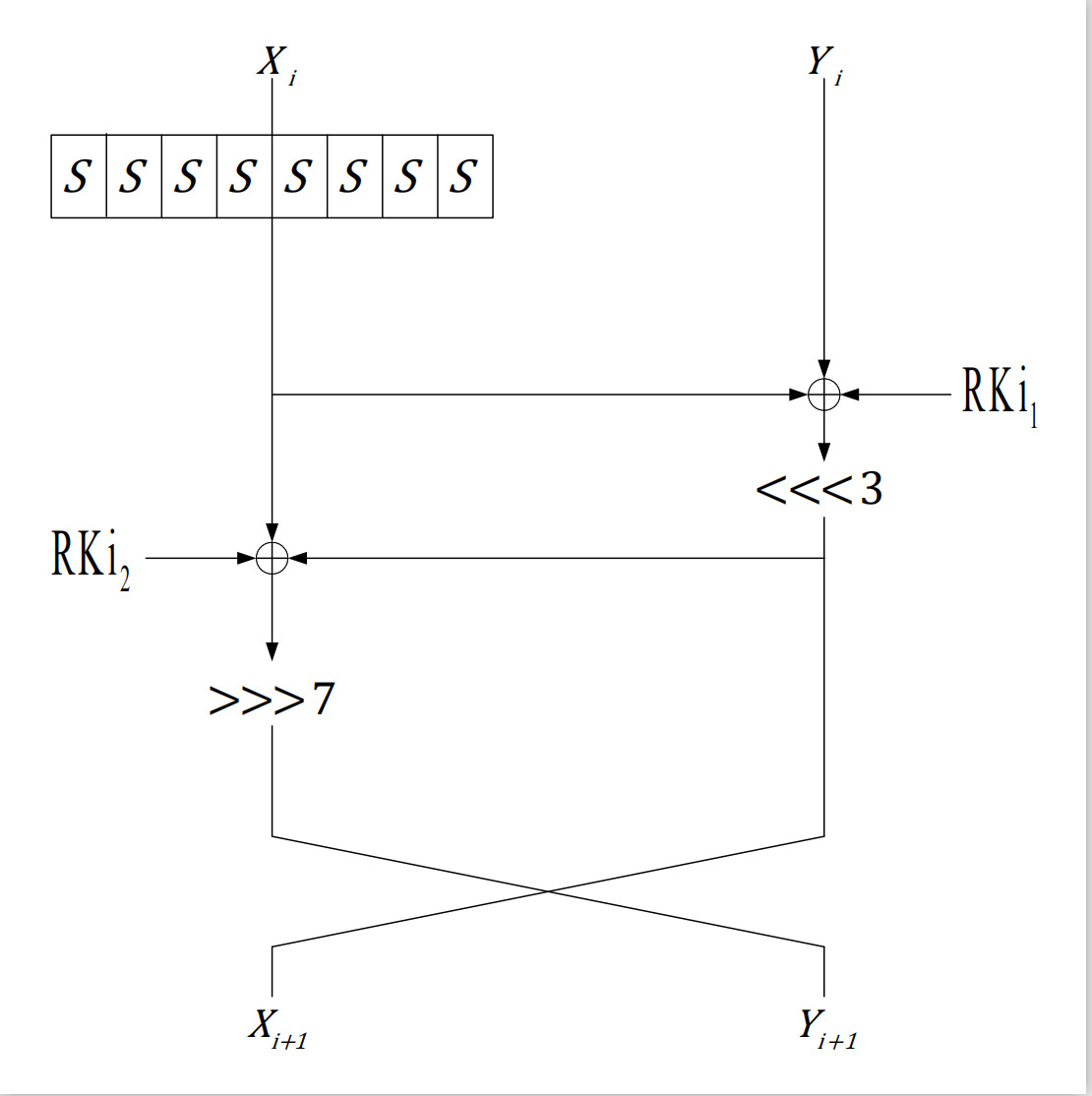}
		\caption{Encryption procedure of LiCi}
		\label{fig:fig6}
	\end{figure}

	As shown in Figure \ref{fig:fig6}, the encryption procedure consists of SubBytes, Cycle shift, and AddRoundKey. The encryption process is denoted as follows:
	
	\begin{align*}\label{eq:equation4}
		X_{i+1} &=[S[X_i]\oplus Y_i \oplus RKi_1]<<<3 
		\\
		Y_{i+1} &=[S[X_i]\oplus X_{i+1}\oplus RKi_2]>>>7
	\end{align*}
	
	The S-box of LiCi is shown in Table \ref{tab:tab2}.
	
	\begin{table*}
		\centering
		\caption{LiCi S-box}
		\label{tab:tab2}
		\begin{tabular}[htbp]{|p{0.2em}|p{0.2em}|p{0.2em}|p{0.2em}|p{0.2em}|p{0.2em}|p{0.2em}|p{0.2em}|p{0.2em}|p{0.2em}|p{0.2em}|p{0.2em}|p{0.2em}|p{0.2em}|p{0.2em}|p{0.2em}|p{0.2em}|}
			\hline
			x&0&1&2&3&4&5&6&8&9&a&b&c&d&e&f \\ \hline
			S&3&f&e&1&0&a&5&c&4&b&2&9&7&6&d\\
			\hline
		\end{tabular}

	\end{table*}
	
	\subsubsection{Key Scheduling}
	
	The key length of LiCi is 128-bit, denoted as $K=K_{127} K_{126}...K_2 K_1 K_0$, where K is a register used to store the input 128-bit key. In each round of encryption, the rightmost 64 bits are taken as the round key, denoted as:
	
	\begin{align*}
		RKi=(RKi_2,RKi_1) \\
		RKi_2=K_{63}K_{62}...K_{34}K_{33}K_{32} \\
		RKi_1=K_{31}K_{30}...K_{2}K_{1}K_{0}
	\end{align*}
	
	The round key will be updated according to the following algorithm:
	
	1.$K<<<13$
	
	2.$[K_3K_2K_1K_0]\gets S[K_3K_2K_1K_0]$
	
	3.$[K_7K_6K_5K_4]\gets S[K_7K_6K_5L_4]$
	
	4.$[K_{63}K_{62}K_{61}K_{60}K_{59}] \gets [K_{63}K_{62}K_{61}]K_{60}K_{59}]\oplus RC^i$
	
	The round counter RCi of 5 bits represents the round i.
	
	\section{Quantum Circuit Implementation of Lblock}
	
	This section proposes a reversible quantum circuit for the LBlock. We first provide the quantum circuit implementation of each component of the round function, then implement the key generation algorithm using quantum circuits. Finally, a complete quantum circuit design scheme is obtained by synthesizing the two parts. For the quantum circuit proposed in this section, only the plaintext and key qubits are allocated without additional qubits. AddRoundKey, S-box, Linear Transformation, and Key Schedule are all optimized from the qubits and quantum gates perspective to ensure minimal resource consumption.

\subsection{Quantum Circuit of Round Function}
	
	\subsubsection{AddRoundKey}
	
	For AddRoundKey, the leftmost 32-bit in the key register is used as the round key $K_i$, round encryption and only 32 CNOT gates are used for each round of its implementation.
	
	\subsubsection{S-box}
	
	For quantum circuit implementation of S-box, there are usually two approaches. One of them is to derive quantum circuits from Algebraic Normal Form (ANF). Here we take $s_0$ as an example, and ANF of $s_0$ is expressed as follows:
	\begin{flalign*}
		&y_0=1+x_0+x_0x_2+x_1x_3+x_2x_3+x_0x_1x_2 \\
		&y_1=1+x_0+x_3+x_0x_1+x_0x_3+x_1x_2+x_1x_3+x_0x_1x_2+\\
		&x_0x_1x_2\\ 
		&y_2=1+x_0+x_1+x_3+x_1x_2+x_1x_3 \\
		&y_3=x_0+x_1+x_2+x_3+x_0x_1
	\end{flalign*}
	
	According to the relationship between variables in the ANF, we can obtain the quantum circuit of the S-box. However, this approach requires a lot of auxiliary qubits and quantum gates. Therefore, we consider using the LIGHTER-R to carry out the in-place implementation of the quantum circuit. The implementation of $s_0$ is shown in Table \ref{tab:tab3} .
	
	\begin{table*}
		\centering
		\caption{In-place implementation of $s_0$}
		\label{tab:tab3}  
		\begin{tabular}[htbp]{cl}%?????????c?????????
			\hline\hline\noalign{\smallskip}
			Algorithms 1:&In-place implementation of $s_0$ \\
			\noalign{\smallskip}\hline\noalign{\smallskip}
			Input:&4-qubit $x_0x_1x_2x_3$ \\
			Output:&4-qubit $y_0y_1y_3y_2$ \\
			1.&$x_3\gets CNOT(x_3,x_0)$\\
			2.&$x_0\gets X(x_0)$\\
			3.&$x_2\gets Toffoli(x_0,x_1,x_2)$\\
			4.&$x_0\gets Toffoli(x_3,x_2,x_0)$\\
			5.&$x_2\gets CNOT(x_2,x_3)$\\
			6.&$x_3\gets X(x_3)$\\
			7.&$x_3\gets Toffoli(x_2,x_1,x_3)$\\
			8.&$x_1\gets Toffoli(x_0,x_3,x_1)$\\
			\noalign{\smallskip}\hline
		\end{tabular}
	\end{table*}
	
\begin{figure}[htb]
		\centering
		\includegraphics[width=4in]{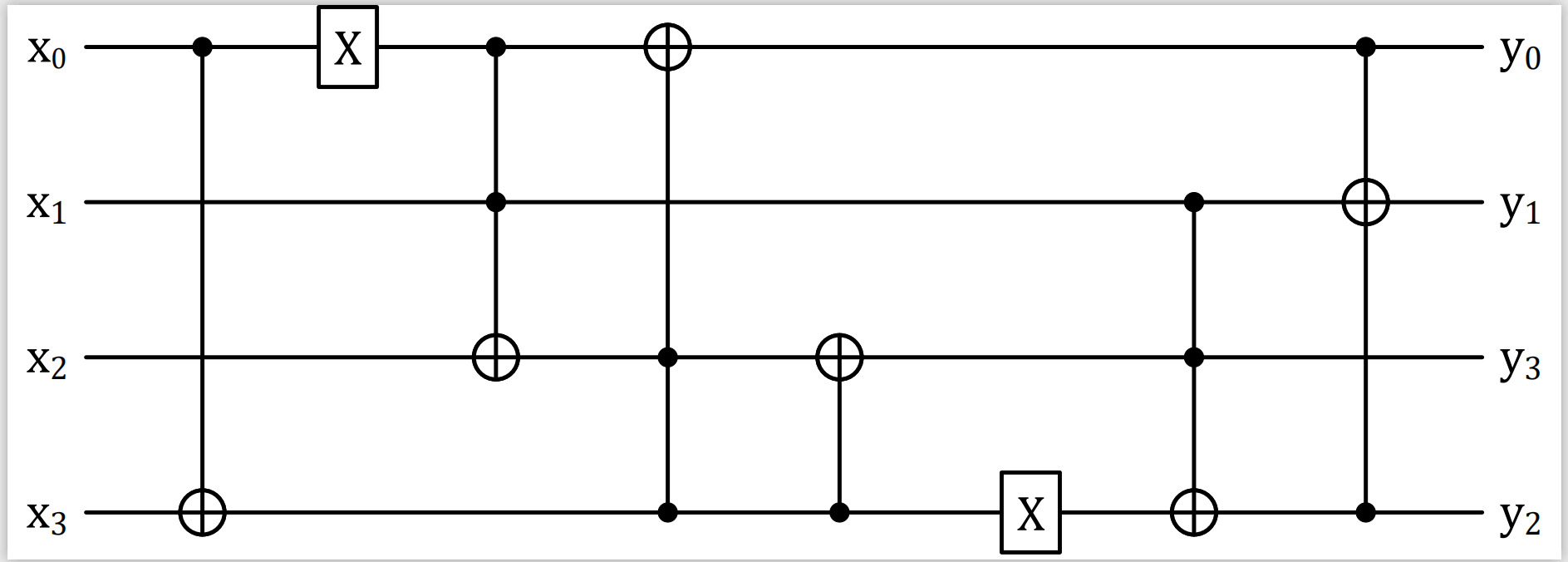}
		\caption{Quantum circuit diagram of $s_0$}
		\label{fig:fig7}
	\end{figure}

	After obtaining the output $y_0y_1y_3y_2$, it needs to be permutated so that it becomes $y_0 y_1 y_2 y_3$. The circuit diagram corresponding to the above instructions one by one is shown in Figure \ref{fig:fig7}.
	
\begin{figure}[!h]
		\centering
		\includegraphics[width=4in]{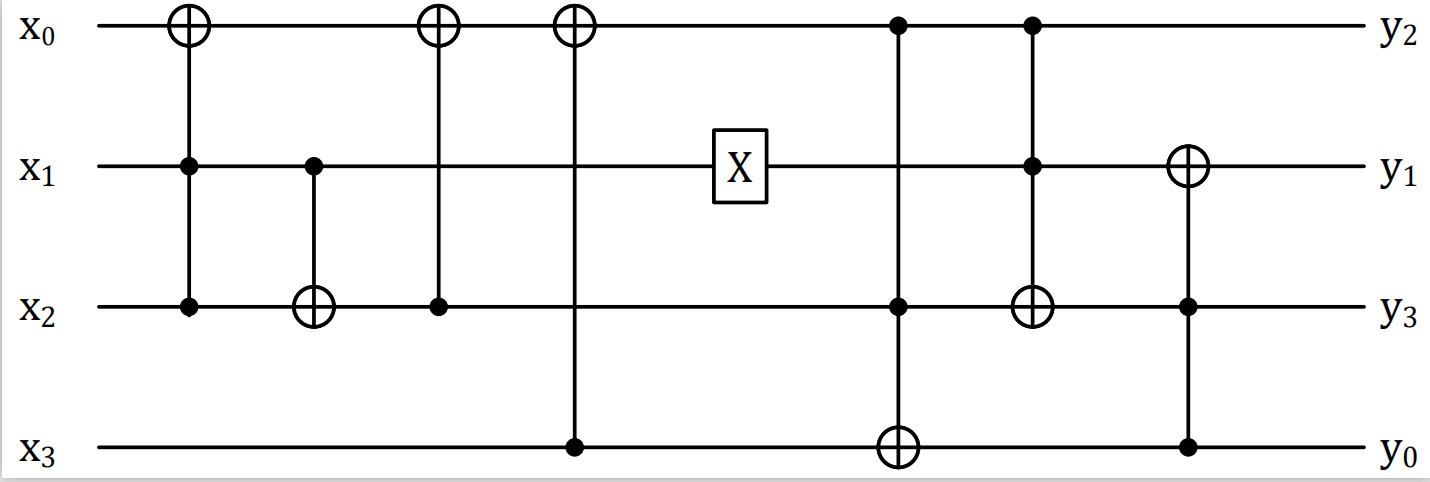}
		\caption{Quantum circuit diagram of $s_1$}
		\label{fig:fig8}
	\end{figure}

\begin{figure}[!h]
		\centering
		\includegraphics[width=4in]{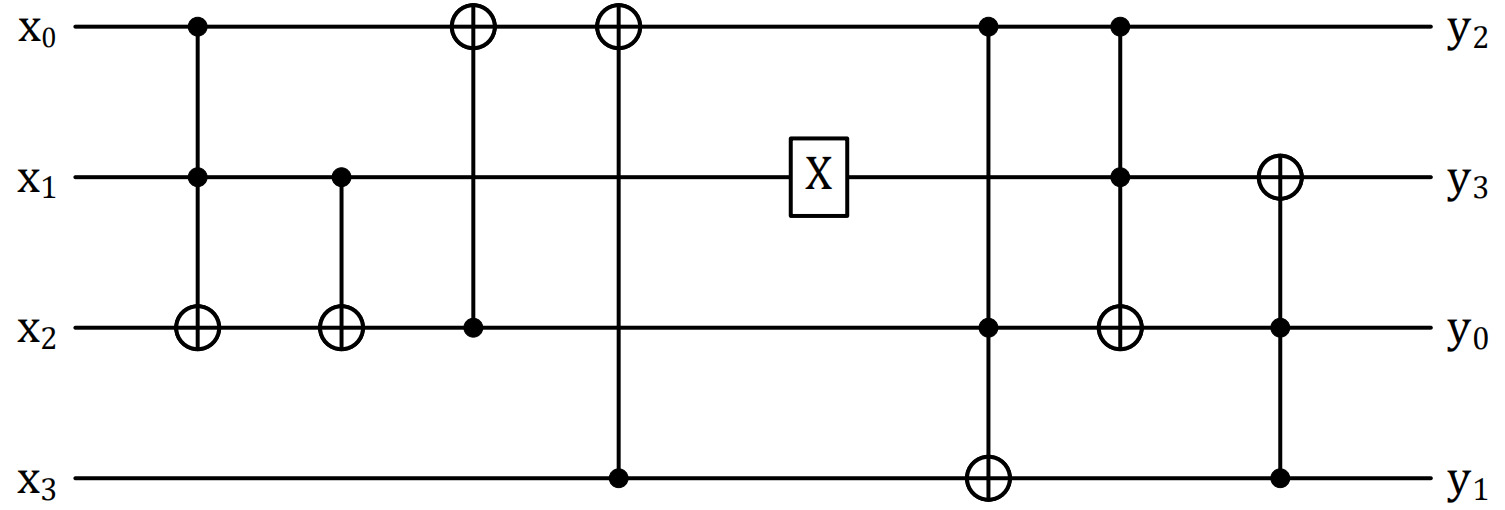}
		\caption{Quantum circuit diagram of $s_2$}
		\label{fig:fig9}
	\end{figure}

\begin{figure}[!h]
		\centering
		\includegraphics[width=4in]{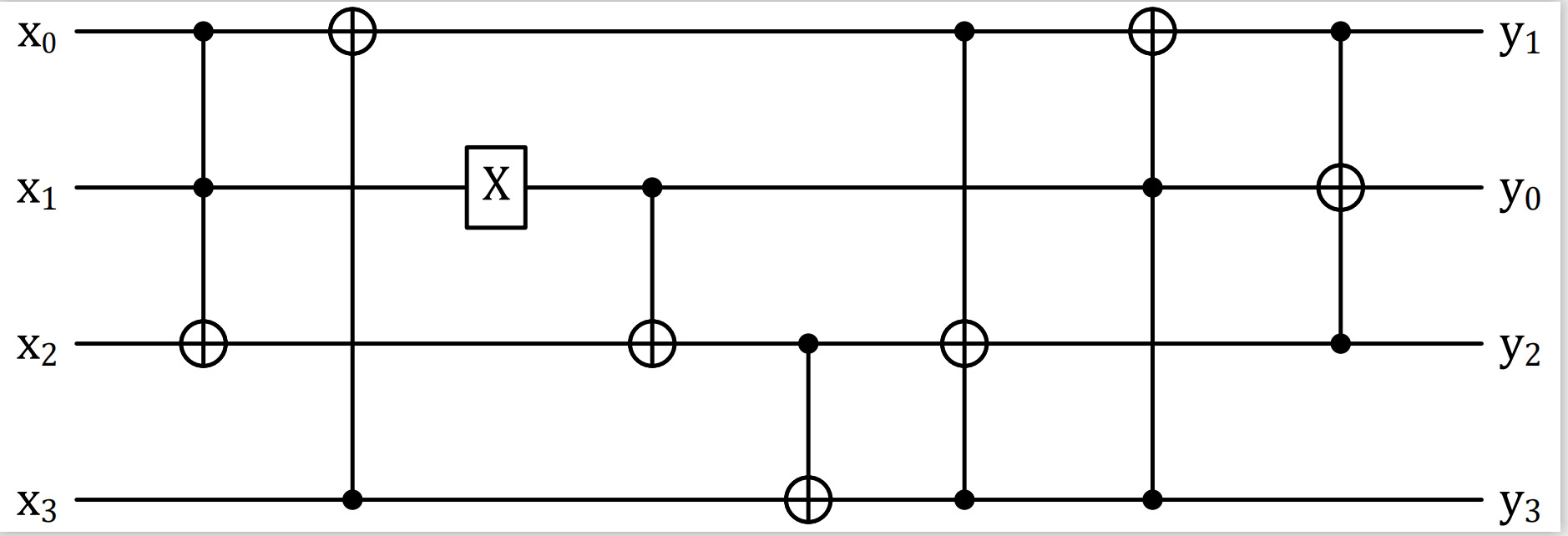}
		\caption{Quantum circuit diagram of $s_3$}
		\label{fig:fig10}
	\end{figure}

\begin{figure}[!h]
		\centering
		\includegraphics[width=4in]{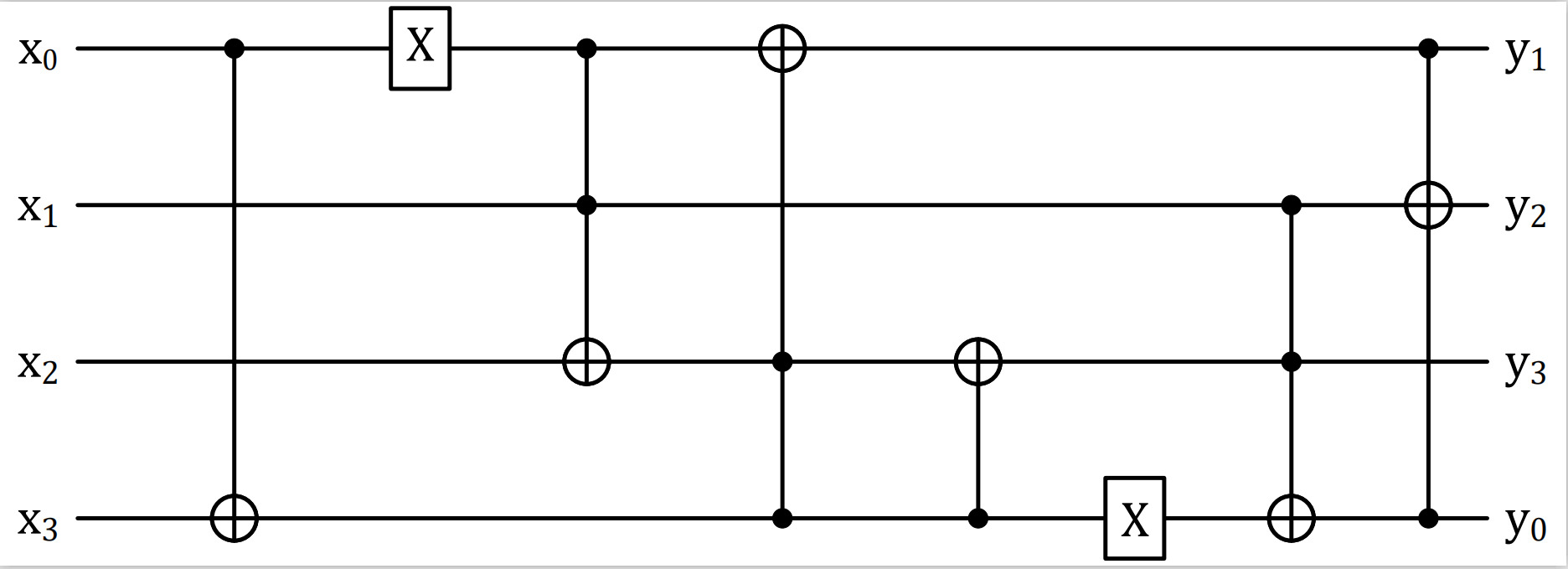}
		\caption{Quantum circuit diagram of $s_4$}
		\label{fig:fig11}
	\end{figure}

\begin{figure}[!h]
		\centering
		\includegraphics[width=4in]{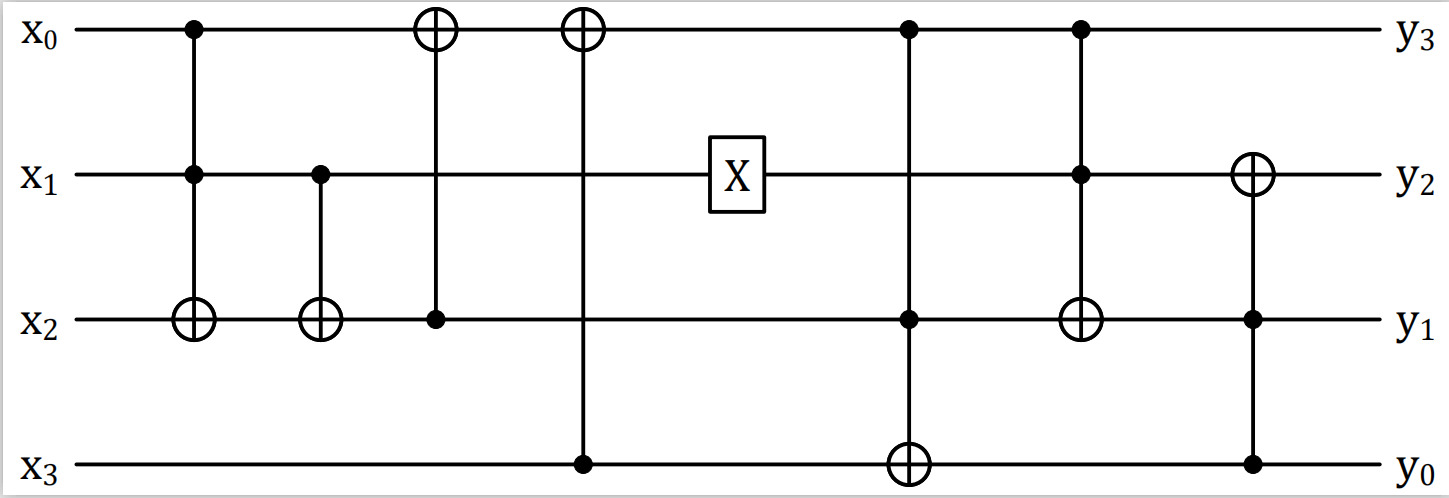}
		\caption{Quantum circuit diagram of $s_5$}
		\label{fig:fig12}
	\end{figure}

\begin{figure}[!h]
		\centering
		\includegraphics[width=4in]{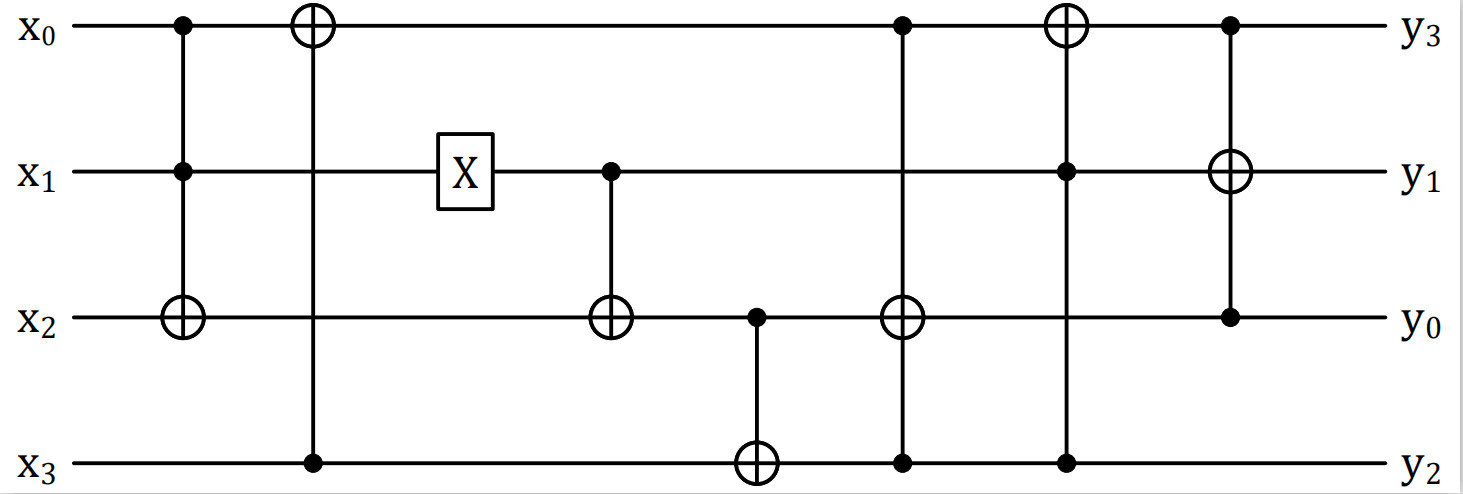}
		\caption{Quantum circuit diagram of $s_6$}
		\label{fig:fig13}
	\end{figure}

\begin{figure}[!h]
		\centering
		\includegraphics[width=4in]{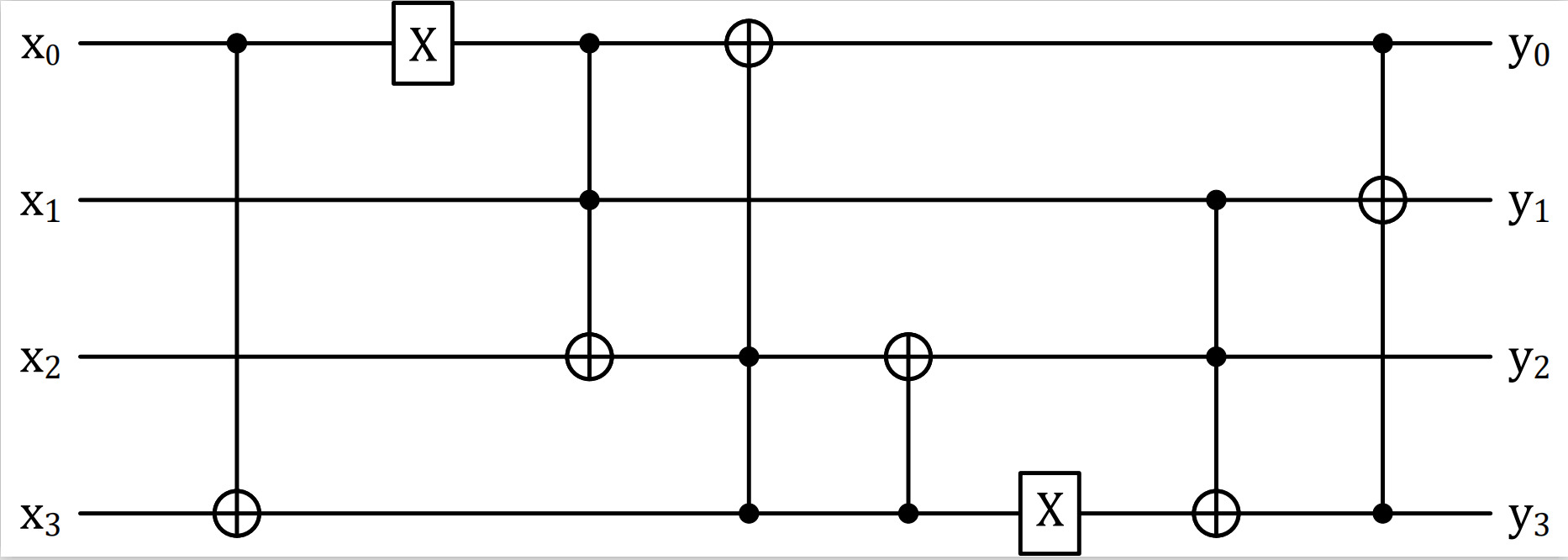}
		\caption{Quantum circuit diagram of $s_7$}
		\label{fig:fig14}
	\end{figure}

	Similarly, we optimized the quantum circuit for other S-boxes $s_1,s_2,s_3,s_4,s_5,s_6$, and $s_7$ of LBlock. The optimized results are shown in Figures \ref{fig:fig8}-\ref{fig:fig14}.
	

\subsubsection{Diffusion Layer and Cyclic Shift}
	After the S-box operation, we perform permutation on the current state. This can be achieved by swapping the positions of the corresponding qubits using quantum SWAP gates. However, since the classical controller can record such an exchange of positions and rearrange its placement in subsequent circuits, SWAP instructions are generally considered quantum resource free. For details, please refer to \cite{bib34}, which was published in EUROCRYPT 2020. Some subsequently published papers on implementing quantum circuits also explain this issue \cite{bib22,bib24,bib26,bib36}. The SWAP operation is also not included in the quantum cost of our work. Similarly, cyclic shifts can also be implemented using quantum SWAP gates to swap the positions of corresponding qubits. This can also be achieved by tracking the reconnection of corresponding qubits. Like the permutation operation, this operation is also considered to consume no quantum resources.

	\subsection{Quantum Circuit of Key Generation Algorithm}
	
\begin{figure}[htbp]
		\centering
		\includegraphics[width=4in]{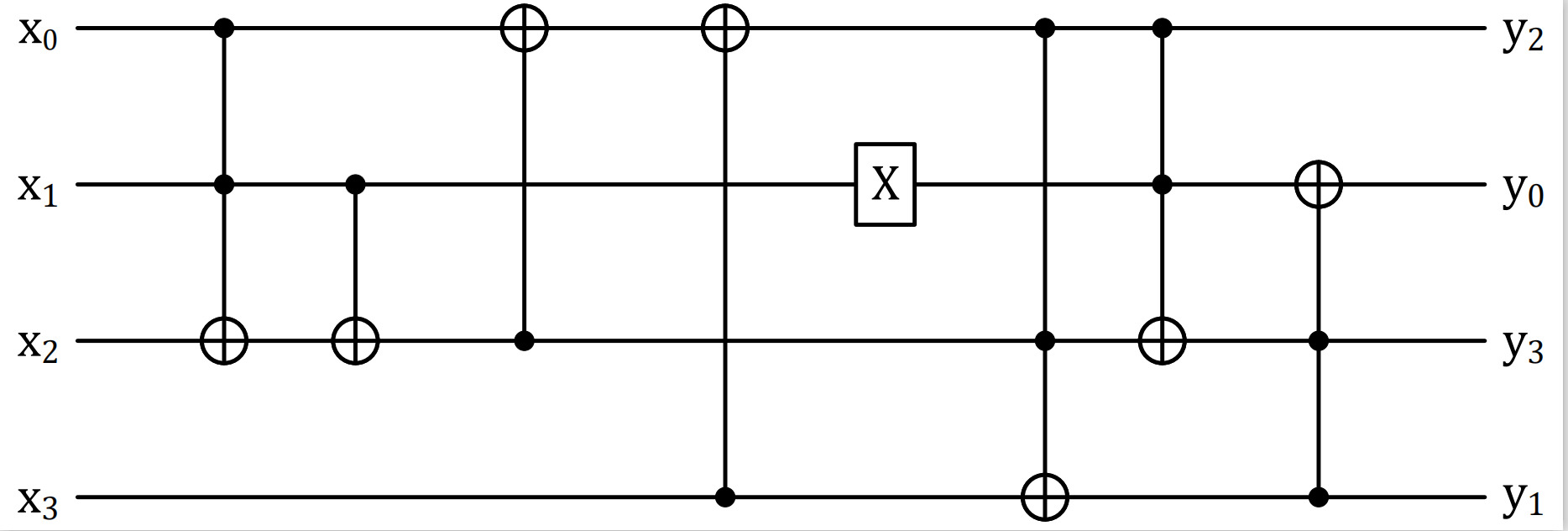}
		\caption{Quantum circuit diagram of $s_8$}
		\label{fig:fig15}
	\end{figure}

\begin{figure}[!h]
		\centering
		\includegraphics[width=4in]{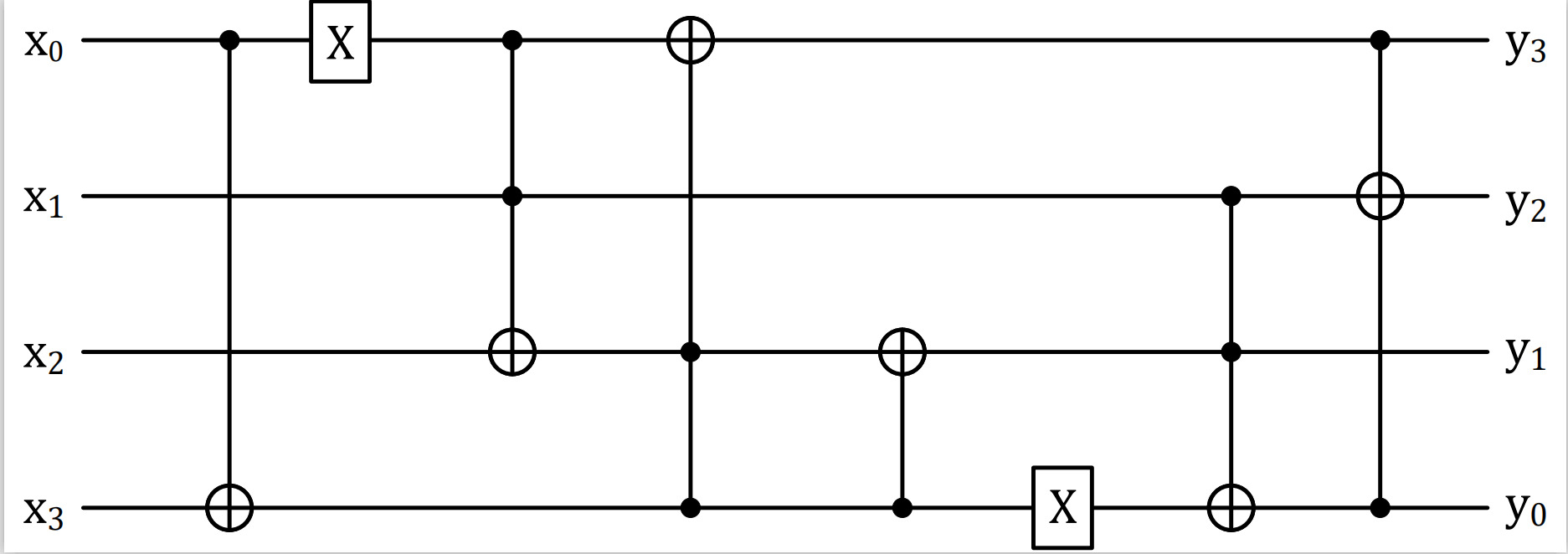}
		\caption{Quantum circuit diagram of $s_9$}
		\label{fig:fig16}
	\end{figure}

	The key expansion algorithm of LBlock includes three parts: cyclic shift, non-linear transformation, and XOR round $i$. By executing the logical exchange of indices between qubits, the cyclic shift can be implemented with no use of quantum resources. After the cyclic shift, the $s_9$ is performed on the leftmost half byte $k_{79} k_{78} k_{77} k_{76}$, and the $s_8$ is performed on the second half byte $k_{75} k_{74} k_{73} k_{72}$. The corresponding quantum circuit diagrams of $s_8$ and $s_9$ are shown in Figure \ref{fig:fig15} and \ref{fig:fig16}. For $[k_{50}k_{49}k_{48}k_{47}k_{46}]\oplus [i]_2$, we can use five CNOT gates for implementation. However, since the number of rounds i is known, we only need to set the NOT gate at the corresponding qubit to invert to complete the XOR of $k_50 k_49 k_48 k_47 k_46$ and $[i]_2$. The key generation algorithm for LBlock is given in Table \ref{tab:tab4}.
	
	\begin{table*}
		\centering
		\caption{Quantum circuits for Keyschedule of Lblock}
		\label{tab:tab4}  
		\begin{tabular}{l}%?????????c?????????
			\hline\hline\noalign{\smallskip}
			Algorithms 2:Quantum circuits for Keyschedule of Lblock \\
			\noalign{\smallskip}\hline\noalign{\smallskip}
			Input:80-qubit key $K(k_{79}, ..., k_0)$ \\
			Output:32-qubit round key $RK(rk_{31}, ..., rk_0)$,\\ 80-qubit update key $K(k_{79}, ..., k_0)$ \\
			1. $k \gets k <<< 29$ using CyclicShift function\\
			2.$k_{79}, k_{78}, k_{77}, k_{76}\gets Sbox(k_{79}, k_{78}, k_{77}, k_{76})$\\
			3.$k_{75}, k_{74}, k_{73}, k_{72}\gets Sbox(k_{75}, k_{74}, k_{73}, k_{72})$\\
			4.$k_{50}, k_{49}, k_{48}, k_{47}, k_{46}\gets $AddConstants $(k_{50}, k_{49}, k_{48}, k_{47}, k_{46})$ \\according to round i\\
			5.return $K(k_{79}, ..., k_{48})$\\
			\noalign{\smallskip}\hline
		\end{tabular}
	\end{table*}
	
	Table \ref{tab:tab5} provides a detailed description of AddConstants
	
	\begin{table*}
		\centering
		\caption{AddConstants}
		\label{tab:tab5}  
		\begin{tabular}{cl}%?????????c?????????
			\hline\hline\noalign{\smallskip}
			Algorithms 3:&AddConstants \\
			\noalign{\smallskip}\hline\noalign{\smallskip}
			Input:&80-qubit key $K(k_{79}, ..., k_0)$,Round constant i \\
			Output:&80-qubit uodate $K(k_{79}, ..., k_0)$,\\ 
			1.&if 1st bit of Round constant = 1 then\\
			2.&$K[46]\gets X(K[46])$\\
			3.&if 2nd bit of Round constant = 1 then\\
			4.&$K[47]\gets X(K[47])$ \\
			5.&if 3rd bit of Round constant = 1 then\\
			6.&$K[48]\gets X(K[48])$\\
			7.&if 4th bit of Round constant = 1 then\\
			8.&$K[49]\gets X(K[49])$\\
			9.&if 5th bit of Round constant = 1 then\\
			10.&$K[50]\gets X(K[50])$\\
			\noalign{\smallskip}\hline
		\end{tabular}
	\end{table*}
	
	\subsection{Overall Quantum Circuit Implementation and Efficiency Improvement}
	
\begin{figure}[!h]
		\centering
		\includegraphics[width=5in]{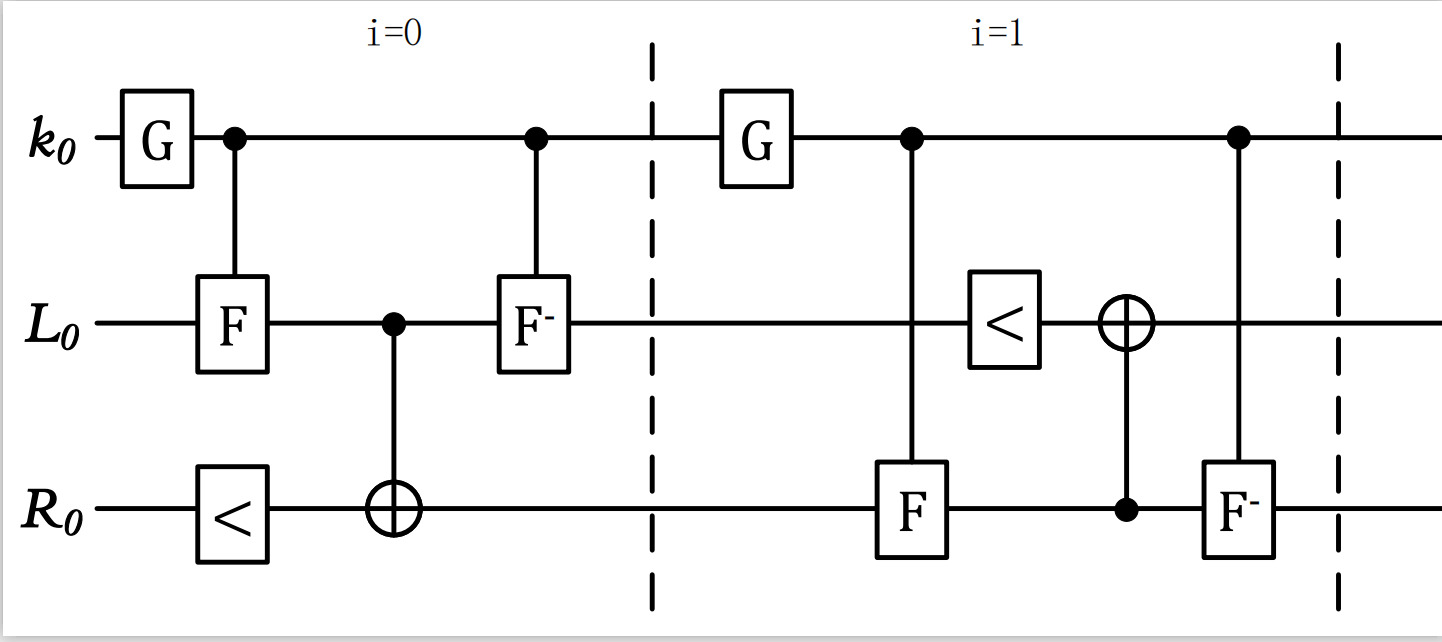}
		\caption{TheoveralldesignoftheLBlockalgorithm}
		\label{fig:fig17}
	\end{figure}

	Based on the analysis of 3.1 and 3.2, we give the overall scheme of the quantum circuit implementation of the LBlock algorithm, as shown in Figure \ref{fig:fig17}.
	
	In the quantum circuit proposed above, after AddRoundKey, S-box and P permutation are performed on $L_i$, for the following operation, the above process needs to be inverted to obtain the data before passing the round function, then the data is used as $R_{(i+1)}$ for the next round function. The inverse operation consumes a lot of resources, so we considered optimizing it.
	
	We observe that the last quantum gates for the 8 S-boxes used in the round function are all Toffoli gates. The Toffoli gate XOR the values of two control qubits to the target qubit. Then, after the transformation of the linear layer, the data is XOR with $R_{(i+1)}$. We also take $s_0$ as an example, in the last Toffoli gate of the circuit, $x_0$ and $x_3$ are the control qubits and $x_1$ is the target qubit. 
	
\begin{figure}[!t]
		\centering
		\includegraphics[width=5in]{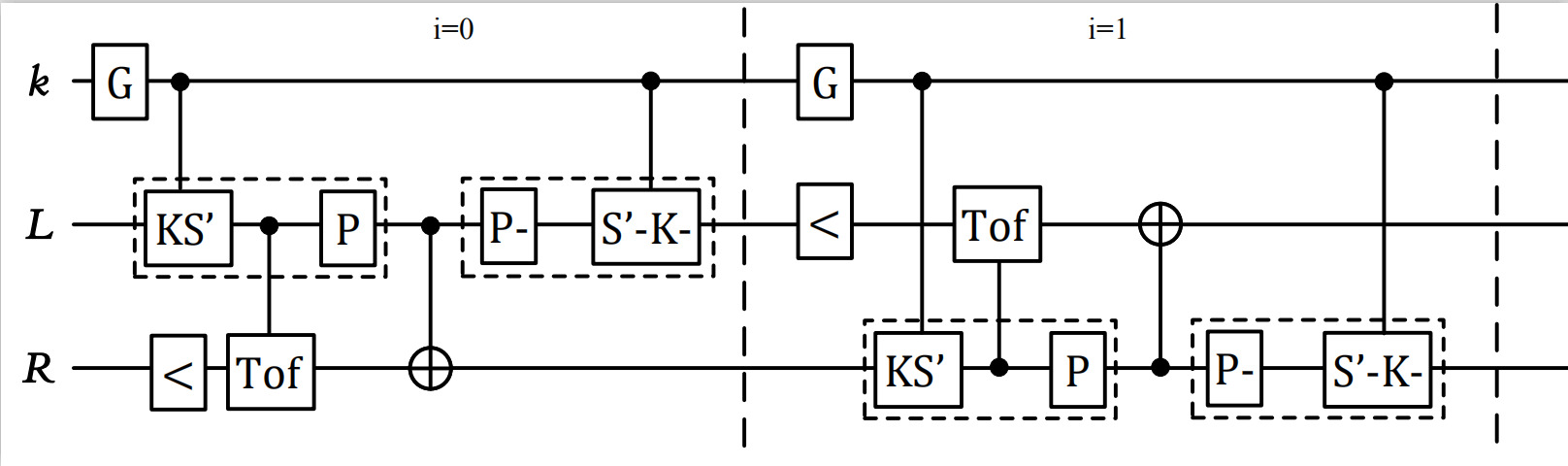}
		\caption{The optimized circuit for the LBlock algorithm}
		\label{fig:fig18}
	\end{figure}

	We can directly XOR the result of the AND operation on $x_0$ and $x_3$ onto the corresponding quantum circuit of $R_i$, so that the Toffoli gate can be reduced once in the process of inverting the data on the left. Since the quantum circuits we design are all reversible, if the implementation of the S-box minimizes the use of the Toffoli gate once, then the Toffoli gate will be reduced once for the related inversion. Based on this idea, we obtain the optimized quantum circuit diagram, as shown in Figure \ref{fig:fig18}.
	
	The specific scheme of LBlock round function optimization is shown in Table \ref{tab:tab6}.
	
	\begin{table*}
		\centering
		\caption{Round function F, Confusion function S and Diffusion function P}
		\label{tab:tab6}  
		\begin{tabular}[htbp]{cl}%?????????c?????????
			\hline\hline\noalign{\smallskip}
			Algorithms 4:&Round function \\
			\noalign{\smallskip}\hline\noalign{\smallskip}
			Input:&32-qubit L, 32-qubit R, 32-qubit k \\
			Output:&32-qubit L, 32-qubit R (after round function)\\ 
			1.&CNOT32(eng, k, L)\\
			2.&L[28:32] = SBOX7(eng, L[28:32])\\
			3.&L[24:28] = SBOX6(eng, L[24:28])\\
			4.&L[20:24] = SBOX5(eng, L[20:24])\\
			5.&L[16:20] = SBOX4(eng, L[16:20])\\
			6.&L[12:16] = SBOX3(eng, L[12:16])\\
			7.&L[8:12] = SBOX2(eng, L[8:12])\\
			8.&L[4:8] = SBOX1(eng, L[4:8])\\
			9.&L[0:4] = SBOX0(eng, L[0:4])\\
			10.&R = S$\_$ plus$\_$b$\_$32(eng, R, 8)\\
			11.&Toffoli$\_$gate(eng, L[1], L[3], R[10])\\
			12.&Toffoli$\_$gate(eng, L[4], L[7], R[2])\\
			13.&Toffoli$\_$gate(eng, L[10], L[11], R[12])\\
			14.&Toffoli$\_$gate(eng, L[13], L[14], R[7])\\
			15.&Toffoli$\_$gate(eng, L[18], L[19], R[25])\\
			16.&Toffoli$\_$gate(eng, L[22], L[23], R[17])\\
			17.&Toffoli$\_$gate(eng, L[24], L[27], R[30])\\
			18.&Toffoli$\_$gate(eng, L[28], L[31], R[22])\\
			19.&L = Permutation(eng, L)\\
			20.&CNOT32(eng, L, R)\\
			\noalign{\smallskip}\hline
		\end{tabular}
	\end{table*}

\section{Quantum Circuit Implementation of LiCi}
	This section proposes a reversible quantum circuit for the LiCi algorithm. Unlike LBlock, LiCi does not require inversion in quantum circuit implementation. In our quantum circuit design, all operations of LiCi, including SubBytes, CyclicShift, AddRoundKey, and key expansion, are optimized from the perspective of minimum quantum resource consumption.
	
	\subsection{Quantum Circuit of Round Function}
	\subsubsection{AddRoundKey}
	In the round function of the LiCi algorithm, the $Rki_2$ and $Rki_1$ generated by the master key are XOR with the left and right branches, respectively. A total of 64 CNOT gates are needed to complete the AddRoundKey.
	\subsubsection{S-box
	}
\begin{figure}[!h]
		\centering
		\includegraphics[width=4in]{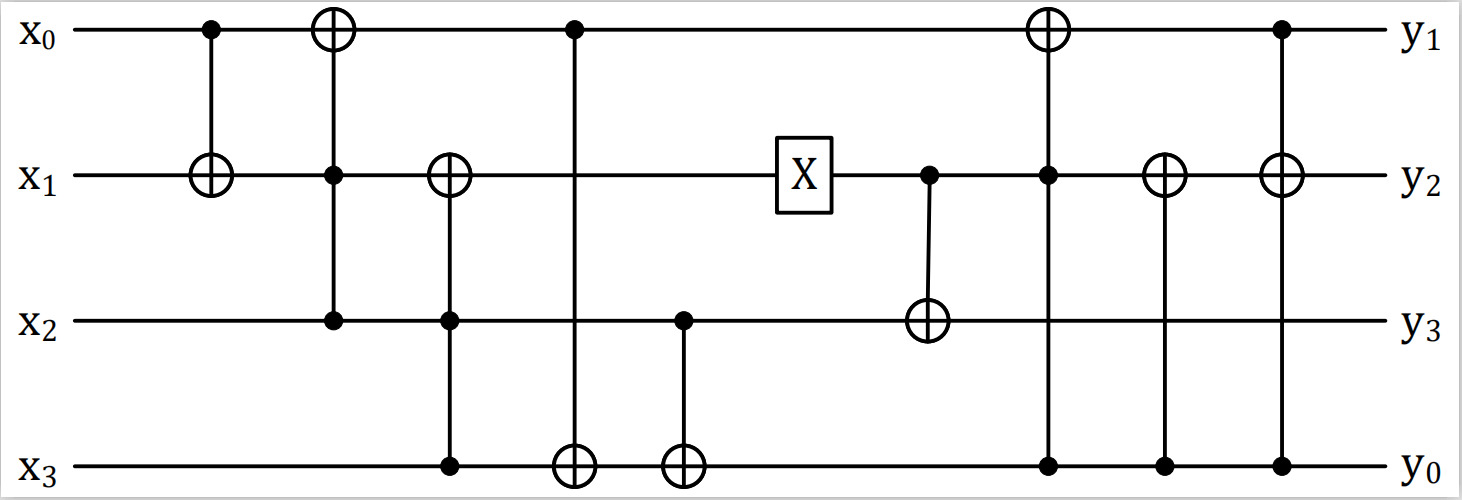}
		\caption{Quantum circuit diagram of s box based on LiCi algorithm}
		\label{fig:fig19}
	\end{figure}

	Since the S-box of LiCi is a lightweight 4-in-4-out S-box, we can use LIGHTER-R to implement it in-place. The realized quantum circuit is shown in Figure \ref{fig:fig19}.
	
	\subsubsection{Cyclic Shift}
	Like Lblocks, cyclic shifts can be achieved using quantum SWAP gates to swap the positions of the corresponding qubits. This can also be achieved by simply tracking the corresponding qubits reconnecting. So, the cyclic shift does not involve the consumption of quantum resources.
	
	\subsection{Quantum Circuit of Key Generation Algorithm}
	The key Scheduling of LiCi is similar to the LBlock; algorithm 5 can be obtained by referring to the design idea of the LBlock key generation algorithm in Section 3.2.
	
	\begin{table*}
		\centering
		\caption{LBlock key generation algorithm}
		\label{tab:tab7}  
		\begin{tabular}[htbp]{cl}%?????????c?????????
			\hline\hline\noalign{\smallskip}
			Algorithms 5:&Lici Key Scheduling \\
			\noalign{\smallskip}\hline\noalign{\smallskip}
			Input:&128-qubit K\\
			Output:&128-qubit K (Key Scheduling)\\ 
			1.&K = S$\_$plus$\_$b$\_$128(eng, K, 13)\\
			2.&K[0:4] = SBOX0(eng, K[0:4])\\
			3.&K[4:8] = SBOX0(eng, K[4:8])\\
			4.&AddConstant(eng, K, round)\\
			\noalign{\smallskip}\hline
		\end{tabular}
	\end{table*}
	
	\subsection{Overall Quantum Circuit Implementation Design for LiCi}
	
\begin{figure}[!h]
		\centering
		\includegraphics[width=5in]{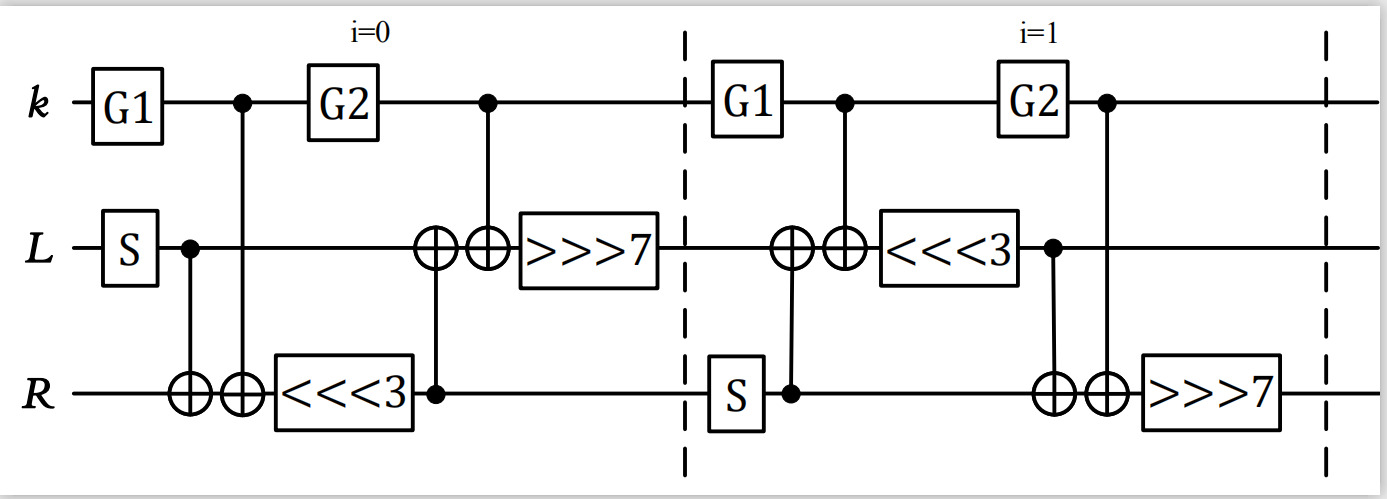}
		\caption{The overall design of the LiCi algorithm}
		\label{fig:fig20}
	\end{figure}

	Based on the analysis of 4.1 and 4.2, we can get the overall design of LiCi, as shown in Figure \ref{fig:fig20}. In each round, the key generation algorithm generates two round keys, which participate in the operation of the right and left branches of the round function, respectively. In the process of the round function, unlike LBlock, the left unit can directly participate in the operation of the right branch without inverting.
	
	\section{Performance Analysis and Quantum Security Evaluation}
	In this section, we analyze the quantum resource consumption of LBlock and LiCi.In past work on the quantum circuit implementation of block cipher, an essential task is to realize the quantum circuit with the least resources. Therefore, it is necessary to conduct a resource analysis on the proposed quantum circuit.
	
	\subsection{Cost of implementing LBlock and LiCi}
	
	We utilized IBM's ProjectQ framework for the evaluation of the proposed quantum circuit. After implementing LBlock and LiCi, we first refer to the plaintext, ciphertext, and key of \cite{bib32} and \cite{bib33} to verify the correctness of the program. Then, the quantum resources are counted using the RecurceCounter provided by the ProjectQ framework. Different from \cite{bib24,bib25}, this resource statistics method is the result of decomposing the Toffoli gate, while \cite{bib24,bib25} is based on the NCT (including NOT, CNOT, Toffoli) quantum gate sets.
	
	Compared with the resources of the two LBlock implementations, the improved implementation reduces 1536 CNOT gates, 512 H gates, 1792 T gates, and the depth of the quantum gate by 73. Compared with the circuit implementations of LBlock and LiCi, due to the fact that LiCi does not perform inversion, the resources of its circuit implementation, except for qubits, are smaller than those of the improved scheme of LBlock.Compared with the circuit implementations of LBlock and LiCi, due to the lack of inverse operation in LiCi, the resource of its circuit implementation is less than that of the improved scheme of LBlock, except for qubits.
	
	\begin{table*}
		\centering
		\caption{Quantum resource consumption of LBlock and LiCi}
		\label{tab:tab8}  
		\begin{tabular}[!h]{ccccccc}%?????????c?????????
			\hline\hline\noalign{\smallskip}
			Algorithms&$\#$Qubits&$\#$CNOT&$\#$H&$\#$T&$\#$X&depth \\
			\noalign{\smallskip}\hline\noalign{\smallskip}
			LBlock(original version)&144&18283&4592&16072&877&1813\\
			LBlock(improved version )&144&16747&4080&14280&877&1740\\ 
			LiCi&192&12900&2464&8624&379&1210\\
			\noalign{\smallskip}\hline
		\end{tabular}
	\end{table*}
	
	To make an intuitive comparison with other quantum circuit implementations of the block cipher in terms of resources, We classify H gates and X gates as 1qCliff gates and compare the quantum resources of LBlock and LiCi with some of the algorithms listed in \cite{bib26}, as shown in Table \ref{tab:tab9}.
	
	\begin{table*}
		\centering
		\caption{quantum resources required for our implementation and other block ciphers}
		\label{tab:tab9}  
		\begin{tabular}[!h]{cccccc}%?????????c?????????
			\hline\hline\noalign{\smallskip}
			Algorithms&$\#$Qubits&$\#$CNOT&$\#$lqClif&$\#$T&$\#$depth \\
			\noalign{\smallskip}\hline\noalign{\smallskip}
			LBlock(improved version )&144&16747&4957&14280&1740\\
			LiCi&192&12900&2843&8624&1210\\ 
			DEFAULT&256&62976&12395&57344&2291\\
			DEFAULT)Another version)&640&76800&13175&62720&2497\\
			GIFT-128/128&256&35840&19377&35840&1520\\
			PRESENT-64/128&128&18230&5628&15624&1179\\
			PIPO-64/128&192&9928&3973&8736&1041\\
			SPECK-128/128&256&73490&15951&55566&36358\\
			LEA-128/128&388&94104&31588&71736&47401\\
			HIGHT-64/128&228&57558&16411&40540&14058\\
			CHAM-128/128&292&58040&14640&34160&37766\\
			\noalign{\smallskip}\hline
		\end{tabular}
	\end{table*}
	
	It can be seen from the comparison that the LBlock and LiCi consume fewer quantum circuit resources than most other lightweight block ciphers that have been implemented.This is because LBlock and LiCi do not have complex encryption components. Except for the S-box, which requires many resources, other parts, such as XOR and cyclic shift, consume fewer quantum resources.

	Based on the quantum circuit implementation proposed in Section 3 and Section 4, we estimate the cost of the exhaustive key search attack for LBlock and LiCi.As mentioned in 2.2, Grover's iteration includes the oracle and diffusion operator. Compared to the oracle, the diffusion operator requires much fewer quantum resources, so the cost of the diffusion operator is ignored \cite{bib18,bib34,bib35}.In this article, we also miss it.
	
\begin{figure*}[!h]
		\centering
		\includegraphics[width=5in]{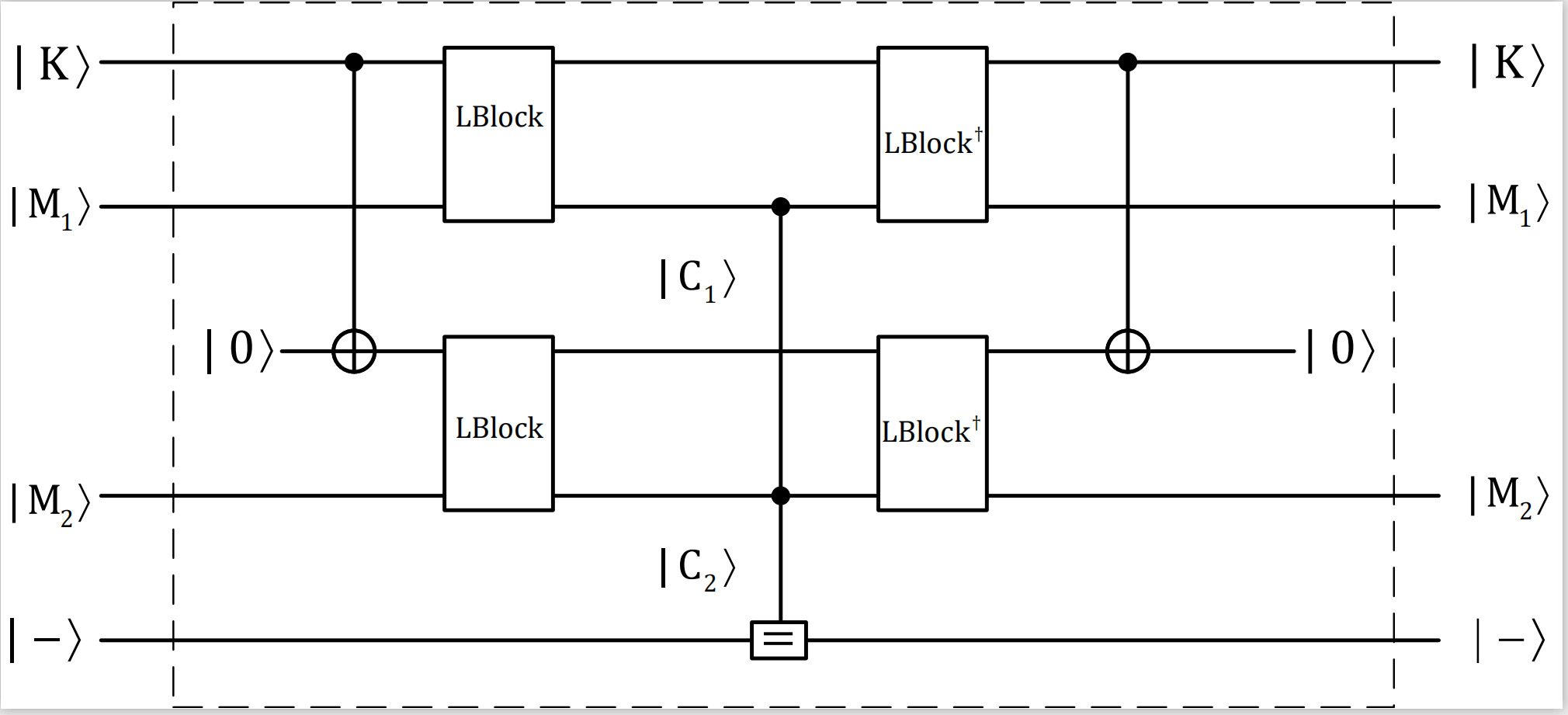}
		\caption{Grover oracle construction from LBlock using two message-ciphertext pairs}
		\label{fig:fig21}
	\end{figure*}

	We first analyze the cost required to build Oracle. For n bit blocks and K bit keys, according to \cite{bib21} and \cite{bib27},$r=\lceil key size/block size\rceil =n/K$ plaintext-ciphertext pairs are used for Grove's key search. For LBlock and LiCi, the corresponding r value is 2. We can construct an oracle for the LBlock algorithm, as shown in Figure \ref{fig:fig21} (LiCi's oracle construction is also similar). The middle operator "=" compares the output of LBlock with the provided ciphertexts and flips the target qubit if they are equal. The cost of performing exhaustive key search attacks is calculated as (Table \ref{tab:tab8})$\times 2\times r\times \lfloor \frac{\pi}{4}\sqrt{N}\rfloor$.
	
	NIST estimates the security strength of symmetric ciphers in the post-quantum era based on the cost of the Grover key search. According to the cost estimates of AES 128, AES 192, and AES 256 by Grassl et al. in the \cite{bib18}, NIST presents the following post-quantum security strengths\cite{bib37}.
	
	Level 1: Any attack that breaks the relevant security definition must require computational resources comparable to or greater than those required for key search on a block cipher with a 128-bit key (e.g. AES-128)
	
	Level 3: Any attack that breaks the relevant security definition must require computational resources comparable to or greater than those required for key search on a block cipher with a 192-bit key (e.g. AES-192)
	
	Level 5: Any attack that breaks the relevant security definition must require computational resources comparable to or greater than those required for key search on a block cipher with a 256-bit key (e.g. AES-256)
	
	The other two levels are level 2 and level 4, which correspond to a 256-bit hash function (e.g. SHA256/ SHA3-256)and a 384-bit hash function(e.g. SHA384/ SHA3-384), respectively.
	
	Through the above analysis, we have calculated the results shown in Table \ref{tab:tab10}.
	
	\begin{table}
		\centering
		\caption{results shown}
		\label{tab:tab10}  
		\begin{tabular}[!h]{cccccc}%?????????c?????????
			\hline\hline\noalign{\smallskip}
			Algorithms&R&$\#Total gates$&$\#$Full depth&$\#$cost&Level of security \\
			\noalign{\smallskip}\hline\noalign{\smallskip}
			LBlock(original version)&2&1.139$\times$ $^{56}$&1.391$\times $2$^{52}$&1.583$\times$2$^{108}$&Not reaching level 1\\
			LBlock(improved version )&2&1.040$\times$2$^{56}$&1.335$\times$2$^{52}$&1.389$\times$2$^{108}$&Not reaching level 1\\ 
			LiCi&2&1.168$\times$2$^{80}$&1.856$\times$2$^{75}$&1.084$\times$2$^{156}$&Not reaching level 1\\
			\noalign{\smallskip}\hline
		\end{tabular}
	\end{table}
	
	We compare the resource consumption of the improved LBlock with that of LiCi. For a single oracle, the cost of quantum gates and circuit depth of LBlock is more than that of LiCi, but the key of LiCi is 128 bits, and the key of LBlock is 80 bits. Therefore, LiCi has much more Grover iterations than LBlock, which lead to more quantum resources for exhaustive key search attack. The key lengths of LiCi and AES-128 are the same, but due to LiCi's lightweight design, its structure is more straightforward, and the cost of constructing Oracle is lower than AES-128. Therefore, the quantum security strength is lower than AES-128, failing to reach Level 1.
	
	It can be predicted that in the post-quantum era, many lightweight block ciphers cannot reach the level 1 security strength specified by NIST. From the quantum security perspective, it is necessary to keep future block ciphers lightweight, while quantum implementation requires as many quantum resources as possible.
	
	In the following work, we will continue to carry out quantum circuit implementation and resource analysis of existing lightweight block cipher, especially block cipher with Feistel structure and ARX structure. At the same time, we are also considering the design of a block cipher that can resist traditional analysis and has high quantum security strength to meet the security needs of the post-quantum era.

\section{Conclusion and Future Work}
	
	In this paper, we propose the quantum circuit implementation of the LBlock algorithm and optimize the overall structure, thereby saving quantum resources and laying the foundation for the efficient implementation of the LBlock algorithm on the quantum computer in the future. At the same time, this design idea provides a reference for the quantum circuit implementation of other Feistel block cipher structures. As a comparison, we also proposed a quantum circuit implementation for LiCi and found that the resource consumption of its circuit implementation is lower than that of LBlock. This is because the quantum implementation process of the LiCi algorithm does not require inversion, which saves a lot of resources. Comparing the quantum circuit implementation of LBlock and LiCi, as well as the implementation of SIMON algorithm in \cite{bib23}, we find that when one branch of the Feistel structure participates in the operation of another branch, if the qubits corresponding to the another branch can be used as a controlled bit to store the calculation results, this implementation does not require inversion and save some quantum resources.
	
	It can be predicted that in the post-quantum era, many lightweight block ciphers cannot reach the level 1 security strength specified by NIST. From the quantum security perspective, it is necessary to keep future block ciphers lightweight, while quantum implementation requires as many quantum resources as possible.
	
	In the following work, we will continue to carry out quantum circuit implementation and resource analysis of existing lightweight block cipher, especially block cipher with Feistel structure and ARX structure. At the same time, we are also considering the design of a block cipher that can resist traditional analysis and has high quantum security strength to meet the security needs of the post-quantum era.

\end{document}